\documentclass{nature}
\usepackage[utf8]{inputenc}
\usepackage{color,ulem}
\usepackage{graphicx}
\usepackage{amssymb}
\usepackage{bm}
\makeatletter
\let\saved@includegraphics\includegraphics
\AtBeginDocument{\let\includegraphics\saved@includegraphics}
\renewenvironment*{figure}{\@float{figure}}{\end@float}
\makeatother
\newcommand{\pasp}{Publications of the Astronomical Society of the Pacific}

\newcommand{\nat}{Nature}
\newcommand{\aap}{Astronomy \& Astrophysic}
\newcommand{\mnras}{MNRAS}
\newcommand{\apj}{Astrophysical Journal} 
\newcommand{\apjl}{Astrophysical Journal Letters}
\newcommand{\apjs}{Astrophysical Journal Supplement}

\newcommand{\prd}{Physical Review D}

\newcommand{\physrep}{Physics Reports}

\begin{document}

\title{A Hubble constant measurement from superluminal motion of the jet in GW170817}



\maketitle

\noindent K. Hotokezaka (Princeton), E. Nakar (Tel Aviv), O. Gottlieb (Tel Aviv), S. Nissanke (GRAPPA University of Amsterdam, Nikhef, Radboud), K. Masuda (NASA Sagan Fellow, Princeton), G. Hallinan (Caltech), K. P. Mooley (Jansky Fellow, NRAO/Caltech), A. T. Deller (Swinburne, OzGrav)
\vspace{1cm}


\begin{abstract}
The Hubble constant ($H_0$) measures the current expansion rate of the Universe, and plays a fundamental role in cosmology.
Tremendous effort has been dedicated over the past decades to measure $H_0$\cite{WMAP,Aubourg2015,Planck2016,Riess2016,Beaton2016,Gao2016,Jang2017,Bonvin2017,Addison2018,DES}.
Notably, {\it Planck} cosmic microwave background (CMB) and the local Cepheid-supernovae distance ladder measurements determine $H_0$ with a precision of $\sim 1$\% and $\sim 2$\% respectively\cite{Planck2016,Riess2016,Riess2018}. A $3$-$\sigma$ level of discrepancy exists between the two measurements\cite{Riess2016,Feeney2018MNRAS}, for reasons that have yet to be understood. Gravitational wave (GW) sources accompanied by electromagnetic (EM) counterparts offer a completely independent standard siren (the GW analogue of an astronomical standard candle) measurement of $H_0$\cite{Schutz1986, Nissanke2010, A17:H0}, as demonstrated following the discovery of the  neutron star merger, GW170817\cite{GW170817,Abbott2017ApJ,Guidorzi2017}. This measurement does not assume a cosmological model and is independent of a cosmic distance ladder. The first joint analysis of the GW signal from GW170817 and its EM localization led to a measurement of $H_0=74^{+16}_{-8}$ km/s/Mpc (median and symmetric $68\%$ credible interval)\cite{A17:H0}. In this analysis, the degeneracy in the GW signal between the source distance and the weakly constrained viewing angle dominated the $H_0$ measurement uncertainty.  Recently, Mooley {\it et al.} (2018)\cite{VLBI} obtained tight constraints on the viewing angle using high angular resolution imaging of the radio counterpart of GW170817. Here we obtain a significantly improved measurement $H_0=68.9^{+4.7}_{-4.6}$ km/s/Mpc 
by using these new  radio observations, combined with the previous GW and EM data. 
We estimate that 15 more localized GW170817-like events (comparable signal-to-noise ratio, favorable orientation), 
having  radio images and light curve data, will potentially bring resolution to the tension between the {\it Planck} and Cepheid-supernova measurements, as compared to 50--100 GW events without such data\cite{Chen2017,Feeney2018}.
\end{abstract}

Mooley {\it et al.} (2018)\cite{VLBI} recently obtained the radio images of a narrowly collimated jet associated with GW170817 by using Very Long Baseline Interferometer (VLBI) and reported the centroid motion of $2.7\pm 0.3$ mas from day 75 to 230, indicating  the  superluminal motion of the jet at an apparent velocity $\beta_{\rm app}=(4.1 \pm 0.4)\left(\frac{d}{41{\rm~Mpc}}\right)$, where $d$ is the source distance from Earth and the velocity is in units of the speed of light, $c$.  In addition, the slow rise\cite{Hallinan2017,mooley2017,Ruan2018} and fast decline\cite{Alexander2018,Margutti2018,VLBI} of the afterglow light curve provide us with evidence that a narrowly collimated jet dominates the emission after the light curve peak. These observations allow us to determine the observing angle independently of the GW analysis.

Given the observed data (the afterglow light curve\cite{Hallinan2017,mooley2017,Dobie2018} at $3$ GHz from day 16 to 294 and the centroid  motion\cite{VLBI}), we constrain the observing angle using several methods: analytic modelling, full hydrodynamic numerical simulations and semi-analytic calculations of synthetic jet models. The analytic modelling and numerical simulations are described in Mooley {\it et al.} (2018)\cite{VLBI}. Giventhe importance of the new constraints on the observing angle on our results, we give here a brief summary of their results. Mooley {\it et al.} (2018) find that the model which best fits the observations is that of a successful jet. We define $\theta_j$ as the jet opening angle, $\theta_{\rm obs}$ as the observing angle, and the difference between them as $\delta_\theta=\theta_{\rm obs}-\theta_j$.  Mooley {\it et al.} (2018)\cite{VLBI} show that the light curve and the small image size imply that the jet must be very narrow, i.e., $\theta_j \ll \delta_\theta$. This implies that the superluminal motion of the jet image can be approximated as that of a point source, where $\delta_\theta \approx 1/\Gamma$ at the time of the observations (near the peak of the light curve). This implies $\delta_\theta \approx 1/\beta_{\rm app} \approx 0.25$ rad and $\theta_j \ll 0.25$ rad, where a source distance of 41 Mpc is assumed. In order to verify this conclusion and to quantify the allowed region for $\delta_\theta$ and $\theta_j \ll 0.25$ rad, they then carried out a set of numerical simulations varying both the opening angle of the jet and the viewing angle allowing for a systematic check of which models can fit both the light curve and the images. They find that only models with  $1/5<\delta_\theta<1/3$  rad and $\theta_j < 0.1$ rad are consistent with observations. They conclude that the combination of the VLBI measurements and the light curve dictates $0.25<\theta_{\rm obs} <0.45~ {\rm rad} ~(15^{\circ}<\theta_{\rm obs}<25^{\circ}$). This constraint is derived assuming that the distance to the source, $d$, is known (41 Mpc). However, in our analysis the distance is unknown and since the main constraint on the observing angle is derived from the the apparent velocity, $\beta_{\rm app} \propto d$, the observing angle is constrained to $0.25<\theta_{\rm obs} \left( \frac{d}{41 {\rm~ Mpc}}\right)<0.45~ {\rm rad}$.

\begin{figure}
\includegraphics[scale=1.1]{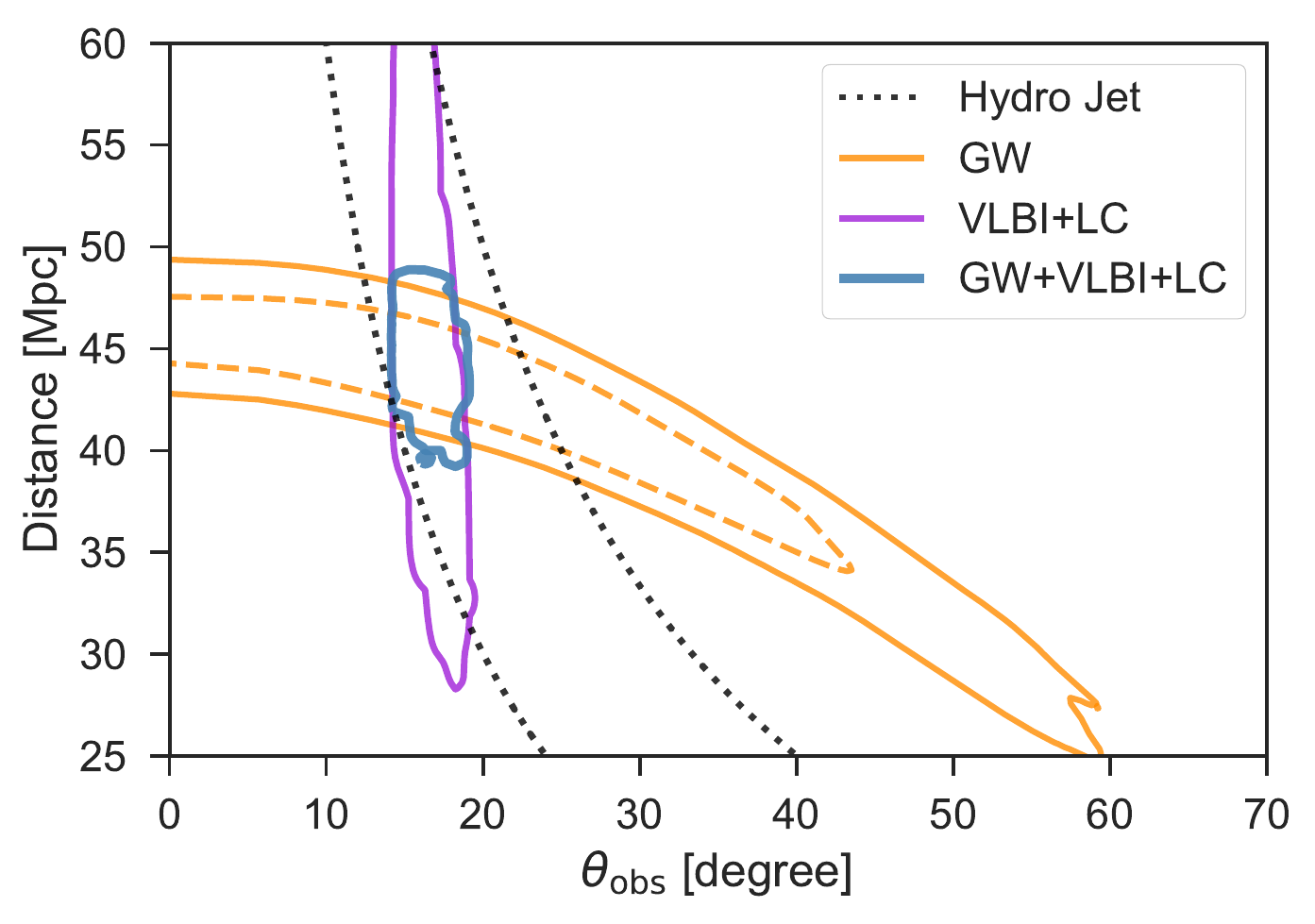}
\caption{
{\bf Distance and observing angle constraints to GW170817.} Dashed curves running from top to bottom depict the constraint of $0.25<\theta_{\rm obs} \left( \frac{d}{41 {\rm~ Mpc}}\right)<0.45~ {\rm rad}$ estimated based on hydrodynamics simulations and synthetic models\cite{VLBI}.
The $95\%$ regions obtained from the MCMC analysis of the afterglow light curve (LC) and centroid motion through Very Long Baseline Interferometry (VLBI) are shown as solid purple (VLBI+LC). The blue contours (VLBI+LC+GW) is the same, but also combined with the GW analysis for a PLJ model. 
Also shown as an orange dashed (solid) contour 
is the $68$ ($95\%$) contour of the posterior distribution of the GW-only analysis (high spin PhenomPNRT posterior samples)\cite{Abbott2018}. We note that the VLBI and light curve data alone provide a distance estimate independent of all other means.}
\label{fig:d-cos}
\end{figure}

In order to obtain the probability distribution of $\theta_{\rm obs}$ and $d$, and to estimate the effect of the jet modelling on the observational constraints on the opening angle, we run also Markov chain Monte Carlo  simulations with two synthetic jet models: a Power-Law Jet (PLJ) and a Gaussian Jet (GJ; see Method). While the hydrodynamics of the jet is not fully taken into account in the synthetic models, unlike the numerical simulations, they allow us to scan the entire parameter space. Therefore, this analysis and the estimate based on the hydrodynamic simulations\cite{VLBI} are complementary.  Figure~\ref{fig:d-cos} shows the posterior distribution for $d$ and $\theta_{\rm obs}$ (see Methods).
The observing angle is constrained to $0.29^{+0.02}_{-0.01}$ rad and $0.30^{+0.02}_{-0.02}$ rad for PLJ and GJ models, respectively.
The constraint on the observing angle for a given model is tighter than the one obtained by the hydrodynamical simulations, most likely because the  simulations explore various outflow structures while each synthetic model explores a single outflow shape. 
The most likely observing angles found with the synthetic models are  smaller by $\sim 0.05$ rad than the median based on the hydrodynamic simulations (but still within the errors). 
We consider this difference as a systematic uncertainty of our analysis (elaborated below), which is most likely attributed to the partial treatment of the hydrodynamic evolution.

\begin{figure}
\includegraphics[scale=0.55]{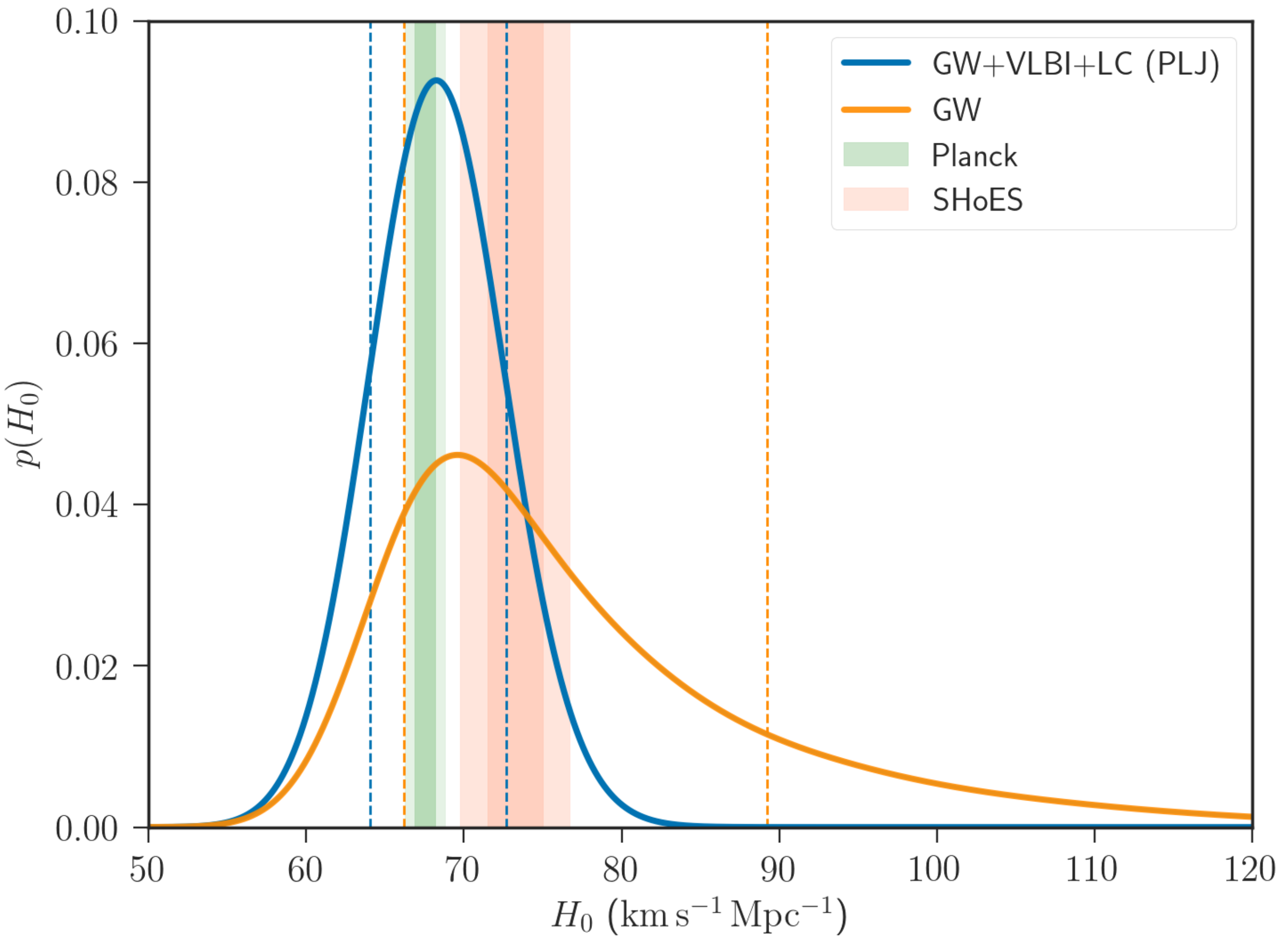}
\caption{
{\bf Posterior distributions for $H_0$.} The results of the GW-only analysis and the combined GW-EM analysis with a PLJ model are shown. The vertical dashed lines show symmetric $68\%$ credible interval for each model. 
The $1$ and $2$-$\sigma$ regions determined by {\it Planck} CMB
(TT,TE,EE+lowP+lensing)\cite{Planck2016} (green) and SH0ES Cepheid-SN distance ladder surveys\cite{Riess2016} (orange) are also depicted as vertical bands. 
}
\label{fig:h0}
\end{figure}

We now turn to the combined GW-EM analysis of the Hubble constant ($H_0$). Namely, we combine the 2-dimensional marginalized GW likelihood distribution (high spin PhenomPNRT)\cite{Abbott2018} for $d$ and $\theta_{\rm obs}$ with that determined from the afterglow light curve and centroid motion (see Methods). 
The posterior distribution for $H_0$ is then computed from the combined likelihood for $d$ and the information about the host galaxy NGC4993 (see Methods)\cite{A17:H0}.
Figure~\ref{fig:h0} depicts the posterior distribution for $H_0$ for a PLJ model and that of the GW-only analysis\cite{A17:H0,Abbott2018}. The constraint is improved from the GW-only analysis, $74^{+16}_{-8}$ km/s/Mpc, to $68.3^{+4.4}_{-4.3}$ km/s/Mpc (median and symmetric $68\%$ credible interval). Also depicted in Figure~\ref{fig:h0} are the regions determined by the {\it Planck}  CMB\cite{Planck2016} and SH0ES Cepheid-supernova distance ladder\cite{Riess2016} surveys respectively. 
Figure 3 shows the posterior distributions for $H_0$ with the different jet models: hydrodynamics simulation jet ($0.25<\theta_{\rm obs} \left( \frac{d}{41 {\rm~ Mpc}}\right)<0.45~ {\rm rad}$), PLJ, and GJ models. The medians and $68\%$ credible intervals are $68.9^{+4.6}_{-4.5}$, $68.3^{+4.4}_{-4.3}$, and $68.5^{+4.4}_{-4.3}$ km/s/Mpc, respectively, corresponding to a precision of 
$6$--$7\%$ at $1$-$\sigma$ level. The sources of errors in our analysis are the GW data, the shape of the light curve, the centroid motion, and the peculiar velocity of the host galaxy. 
While the constraint on $\theta_{\rm obs}$ is slightly different between the three models, the systematic error in $H_0$ due to this difference is much smaller than $7\%$. 
This is because the uncertainty in $H_0$ of our analysis is dominated by both the GW data and the peculiar motion of NGC 4993 (contrary to the GW-only analysis, where the uncertainty in the observing angle is a major source of error).  
Finally, it is important to bear in mind that our result does not depend on the spin prior in the GW analysis\cite{Abbott2018} (see Methods).

Our new analysis, which is based on this single event, improves the $H_0$ measurement to a precision of $\sim 7\%$.  We expect that the precision of the measurement will improve by observing more merger events similar to GW170817, i.e, mergers with detectable jet afterglows. In the coming years, several to tens of neutron star binary mergers (including neutron star-black hole binary systems) per year may be observable in GWs as the LIGO and Virgo detectors improve their sensitivity due to instrument upgrades, and as additional detectors join the GW network\cite{2018LRR....21....3A}. 
In addition, radio afterglow fluxes of merger events at further distances are not necessarily fainter than GW170817 because of the wide variation in the circum-merger densities. For instance, the superluminal motion of a jet can be measured for events taking place out to $\sim 100$ Mpc if the density is about the typical value inferred from short GRB observations\cite{Fong2015} (and the other afterglow parameters are assumed to be the same as GW170817). We note however that a favorable viewing angle is a likely prerequisite for detection. 
For events at greater $d$, while the error due to the radio observations increases, the error due to the peculiar motion decreases. Furthermore, inferring the binary inclination from GW-alone relies on the measurement of the GW polarization, which was particularly challenging in the case of GW170817 because of the low signal-to-noise ratio in the Virgo detector and the two LIGO detectors being nearly co-aligned\cite{GW170817,A17:H0}. For future GW radio jet events with similar signal-to-noise, the $H_0$ uncertainty would thus remain comparable or better to that of this analysis because of the addition of GW detectors and of improved instrument sensitivity\cite{arXiv1307.2638N,Feeney2018,Chen2017}. To achieve a measurement of $H_0$ with  a high  precision using more events, the systematic uncertainty resulting from jet modeling should also be reduced.

Most current methods to estimate $H_0$  span from the local Universe to the CMB and include the use of Cepheid variables and red-giant stars\cite{Jang2017}, supernovae (SNe)\cite{Beaton2016,Riess2016,Riess2018}, circumnuclear megamasers\cite{Gao2016}, gravitational lenses\cite{Bonvin2017}, galaxies\cite{Aubourg2015,Addison2018,DES} and the CMB\cite{WMAP,Planck2016}. These methods either depend on a cosmic distance ladder relating geometric distances of Cepheid variables to standard candles, such as Type 1a supernovae, or assume a certain cosmological model, such as $\Lambda$-CDM\cite{WMAP,Aubourg2015,Planck2016,Riess2016,Beaton2016,Jang2017,Bonvin2017,Addison2018,DES}. The use of geometric distances to circumnuclear megamasers is a notable exception, but is currently limited to $6\%$ precision\cite{Gao2016}.  The current $\gtrsim 3\sigma$ discrepancy\cite{Riess2016,Feeney2018MNRAS} between {\it Planck} CMB measurements and SH0ES  data is of particular interest given the degree of precision in both measurements and the possible implication of the requirement of new physics beyond $\Lambda$-CDM models if the discrepancy turns out to be true (rather then a result of systematic errors)\cite{Weinberg2013}. {\it Gaia} DR2 data on Galactic Cepheids, together with dedicated {\it HST} observations on the latter sample, will likely reduce systematic uncertainties sufficiently to improve the standard candle/distance ladder measurements of $H_0$ to $\sim 1\%$ precision within the next few years\cite{Riess2018}, potentially raising this discrepancy above $5\sigma$. 
A standard siren based  measurement of $H_0$, on a similar timescale, would be particularly useful, as it would independently provide a local measurement of $H_0$ that does not rely on a cosmic distance ladder, and which does not assume any cosmological model as a prior (although there are model assumptions in the interpretation of the VLBI data). We estimate that, after observing $\sim 15$ more GW170817-like events with VLBI data and light curve (comparable SNR, favorable orientation),
as compared to $\sim 50$--$100$ GW events without such data, the precision of the $H_0$ measurement would be $\sim 1.8\%$\cite{arXiv1307.2638N,Feeney2018,Chen2017}. Thus, joint GW-VLBI constraints on $H_0$ will potentially resolve the current tension between {\it Planck} and standard candle/distance ladder data. 

\clearpage

\begin{figure} 
\includegraphics[scale=0.55]{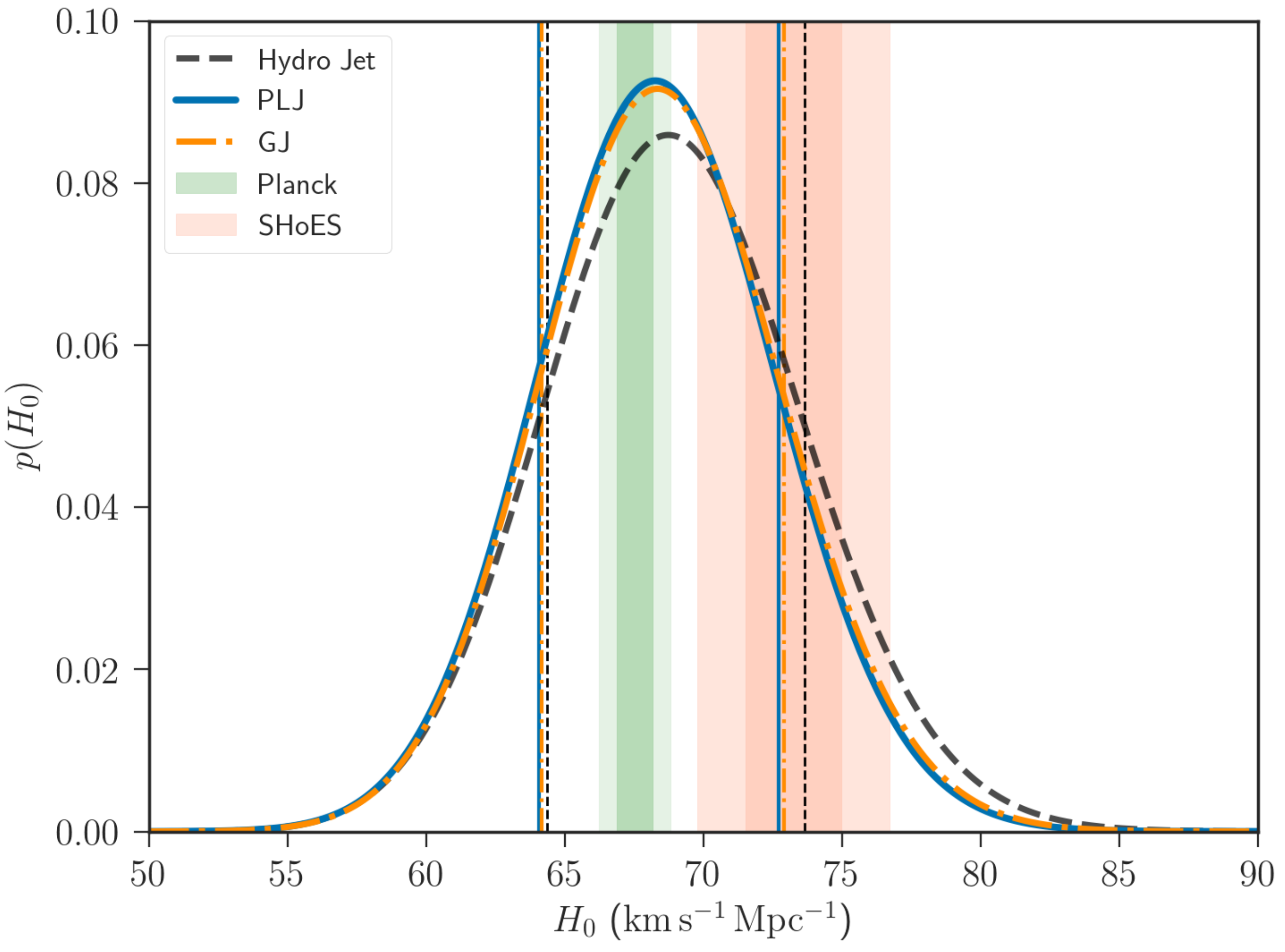}
\caption{
{\bf The Hubble constant with  different jet models.} 
Dashed curve: hydrodynamics simulation jet ($0.25<\theta_{\rm obs} \left( \frac{d}{41 {\rm~ Mpc}}\right)<0.45~ {\rm rad}$), solid curve: a Power-Law Jet, and dash-dotted curve: a Gaussian Jet. 
The vertical lines show symmetric $68\%$ credible interval for each model. 
The $1$ and $2$-$\sigma$ regions determined by {\it Planck}  CMB
(TT,TE,EE+lowP+lensing)\cite{Planck2016} (green) and SH0ES Cepheid-SN distance ladder surveys\cite{Riess2016} (orange) are also depicted as vertical bands. 
}
\label{fig:h0_comp}
\end{figure}

\clearpage

\begin{methods}

\subsection{Light curve and centroid motion modeling}

In the case of the afterglow of GW170817, the observed light curve rules out the simple top-hat jet model and support structured jet models\cite{Hallinan2017,mooley2017,Avanzo2018,Ruan2018,Margutti2018,Troja2018,Lamb2018,Lyman2018,Gill2018,Resmi2018,Nakar2018,2018arXiv180409345X}, of which the structure is likely composed of the jet core and surrounding cocoon\cite{VLBI,Gottlieb2018,Lazzati2017,2018arXiv180409345X}.
We use two different structured jet models: (1) a Power-Law Jet (PLJ)  and (2) a Gaussian Jet (GJ) model, which can mimic the jet-cocoon structure obtained from numerical simulations\cite{VLBI,Gottlieb2018,Lazzati2017,2018arXiv180409345X}.   The isotropic-equivalent energy and initial Lorentz factor vary with the polar angle for a PLJ model:
\begin{eqnarray}
E_{\rm iso}(\theta) & = & \frac{E_{{\rm iso},c}}{1+(\theta/\theta_c)^{\alpha_E}},\\
\Gamma_i (\theta) & = & 1 + \frac{\Gamma_{i,c}}{1+(\theta/\theta_c)^{\alpha_g}},
\end{eqnarray}
where $E_{{\rm iso},c}$, $\theta_{c}$, $\alpha_E$, and $\alpha_g$ are 
free parameters and we fix $\Gamma_{i,c}$ to be $600$. 
For a GJ model:
\begin{eqnarray}
E_{\rm iso}(\theta) & = & E_{{\rm iso},c}\exp\left[- \frac{1}{2} \left(\frac{\theta}{\theta_c}\right)^2\right],\\
\Gamma_i (\theta) & = & 1 + (\Gamma_{i,c}-1)\exp\left[- \frac{1}{2} \left(\frac{\theta}{\theta_c}\right)^2\right],
\end{eqnarray}
where $E_{{\rm iso},c}$,  $\theta_{c}$ are free parameters and we fix $\Gamma_{i,c}$ to be $100$.

For a given set of the model parameters and circum-merger density, $n$, we evolve the jet adiabatically and neglect the lateral expansion\cite{Gill2018}. This assumption is  valid until  the jet slows down sufficiently. For the core of the jet, the lateral expansion occurs on a time scale much longer than what we have considered here, and indeed, we find lack of significant lateral expansion also the  hydrodynamical simulations\cite{VLBI}. For the wing of the jet, however, the lateral expansion is important on the time scales considered here\cite{VLBI}. Therefore, our approximation here is expected to slightly underestimate the observing angle.

 Given a jet evolution, we calculate the afterglow light curve and the motion of the flux center by using the standard synchrotron afterglow model\cite{Sari1998}. The code is described in Hotokezaka and Piran (2015)\cite{Hotokezaka2015}.
 In the case of GW170817, the afterglow has a single power-law spectrum with a spectral index of $0.588\pm 0.005$ from radio to X-ray band\cite{Troja2018,Margutti2018,Alexander2018}, which is consistent with optically thin synchrotron emission in the slow cooling regime. Thus, here we consider only this regime. The synchrotron modeling involves three microphysics parameters ($p,\, \epsilon_e,\, \epsilon_b$), where  $\epsilon_e$ and $\epsilon_b$ are the conversion efficiency from the internal energy to the energy of accelerated electrons and magnetic field, and $p$ is the power-law index of the number distribution of accelerated electrons. Since the power-law index, p, is related to the observed spectrum as $F_{\nu} \propto \nu^{-(p-1)/2}$, we adopt $p=2.16$. We also fix $\epsilon_e$ to be $0.1$.


Assuming the above models, we run Markov-Chain Monte Carlo (MCMC) simulations by using an open code \texttt{emcee}\cite{emcee}.
For the modelling, we use $E_{{\rm iso},c}/n$, which determines the deceleration time scale of the jet, instead of using $E_{{\rm iso},c}$ and $n$ separately to reduce the number of free parameters. Furthermore, instead of using $\epsilon_b$, we introduce an auxiliary parameter, $e_b$, which controls the overall amplitude of the light curve.  
Therefore, in total, we have 7 parameters ($E_{{\rm iso},c}/n$, $\theta_c$, $\alpha_E$, $\alpha_g$, $e_b$, $\theta_{\rm obs}$, $d$) for  PLJ model and 5 parameters ($E_{{\rm iso},c}/n$, $\theta_c$, $e_b$, $\theta_{\rm obs}$, $d$) for  GJ model. 
We adopt a log flat prior for $E_{{\rm iso},c}/n$ and $e_b$, and uniform prior for $\theta_c$, $\alpha_E$, $\alpha_g$, an isotropic prior for $\theta_{\rm obs}$, and a volumetric prior for $d$. 

Figure \ref{fig:d-cos} (VLBI+LC) shows the resulting posterior for $d$ and $\theta_{\rm obs}$ marginalized over the other model parameters. The corner plots for the model parameters are shown in Extended Data Figures \ref{fig:PLJ} and \ref{fig:GJ}. 

\subsection{Combined GW-EM analysis of the Hubble constant}

Next we perform the modeling of the light curve $x_{\rm LC}$ and centroid motion data $x_{\rm VLBI}$, taking into account the constraint from the GW data $x_{\rm GW}$. Because the GW and EM data are independent and only $d$ and $\theta_{\rm obs}$ in the GW model affect the EM data, this can be done by replacing the prior on $d$ and $\theta_{\rm obs}$ in the above MCMC analysis with the marginal posterior distribution from the GW analysis, $p(d, \theta_{\rm obs}|x_{\rm GW})$. Figure \ref{fig:d-cos} (GW+VLBI+LC) shows the resulting posterior distribution $p(d, \theta_{\rm obs}|x_{\rm GW}, x_{\rm VLBI}, x_{\rm LC})$ marginalized over the other model parameters. The corresponding corner plots for the model parameters are shown in Extended Data Figures \ref{fig:PLJGW} and \ref{fig:GJGW}. The posterior models for the afterglow flux at 3 GHz and centroid motion from day 75 to 230 measured with VLBI \cite{Hallinan2017,mooley2017,VLBI} are shown in Extended Data Figure \ref{fig:exp} with the data. 

We combine $p(d, \theta_{\rm obs}|x_{\rm GW}, x_{\rm VLBI}, x_{\rm LC})$ from this joint modeling with the recessional velocity $v_{\rm r}$ to derive the Hubble constant $H_0$. 
To do so, one needs to take into account the unknown peculiar velocity of NGC 4993 as $v_{\rm r}=H_0d+v_{\rm p}$. Here we follow the procedure used in Abbott et al. (2017) \cite{A17:H0} to compute the marginalized posterior for $H_0$:
\begin{eqnarray}
\nonumber
 &&p(H_0|x_{\rm GW}, x_{\rm VLBI}, x_{\rm LC}, v_{\rm r}, \langle v_{\rm p}\rangle)\\
 \nonumber
 &&=\int\mathrm{d}d\,\mathrm{d}\cos\theta_{\rm obs}\,\mathrm{d}v_{\rm p} \, p(H_0, d, \cos\theta_{\rm obs}, v_{\rm p}|x_{\rm GW}, x_{\rm VLBI}, x_{\rm LC}, v_{\rm r}, \langle v_{\rm p}\rangle) \\ 
 && \propto  p(H_0)\int\mathrm{d} d\,\mathrm{d}v_{\rm p} \, p(v_{\rm r}|d, v_{\rm p}, H_0) \,p(\langle v_{\rm p}\rangle|v_{\rm p}) \,p(v_{\rm p})\,
	p(d|x_{\rm GW}, x_{\rm VLBI}, x_{\rm LC}).
\end{eqnarray}
We adopt the same information on $v_{\rm r}$ and $\langle v_{\rm p} \rangle$ as in (Abbott et al 2017)\cite{A17:H0}:
\begin{eqnarray}
p(v_{\rm r}|d,v_{\rm p},H_0) & =& \frac{1}{\sqrt{2\pi \sigma_{v_r}^2}}\exp\left[-\frac{1}{2}\left(\frac{v_{\rm r}-v_{\rm p}-H_0d}{\sigma_{v_r}}\right)^2 \right],\\
p(\langle v_{\rm p} \rangle|v_{\rm p}) & =& \frac{1}{\sqrt{2\pi \sigma_{v_p}^2}}\exp\left[-\frac{1}{2}\left(\frac{\langle v_{\rm p}\rangle-v_{\rm p}}{\sigma_{v_p}}\right)^2 \right],
\end{eqnarray}
where  $v_{\rm r}=3327$ km/s, $\sigma_{v_r}=72$ km/s, $\langle v_{\rm p}\rangle=310$ km/s, and $\sigma_{v_p}=150$ km/s.

The posterior distribution for $H_0$ generally depends on the prior in the GW analysis\cite{Abbott2018}, i.e., the high or low spin  prior. Figure \ref{fig:low} compares the $H_0$ posterior of the high spin prior with 
that of the low spin prior\cite{Abbott2018}. In the case of the GW-only analysis, they depend on the prior as $78^{+20}_{-10}$ km/s/Mpc (low spin) and $74^{+15}_{-8}$ km/s/Mpc (high spin). However, in the case of the combined analysis, they result in practically the same $H_0$, $68.9^{+4.6}_{-4.5}$ km/s/Mpc.
We also did the same analysis by using the GW posterior data of Finstad et al. (2018)\cite{Finstad2018}. These result in slightly smaller values of $H_0$ compared to those with Abbott et al (2018)\cite{Abbott2018}. Note also that our result is consistent with $H_0=71.9\pm7.1$ km/s/Mpc measured by using  the surface brightness fluctuation method applied to NGC 4993\cite{Cantiello2018}, which is calibrated with the  Cepheid  distance measurements.

\clearpage
\setcounter{figure}{0}  

\section*{Extended Data}
\begin{figure}
\hspace{0.5cm}
\includegraphics[scale=0.75]{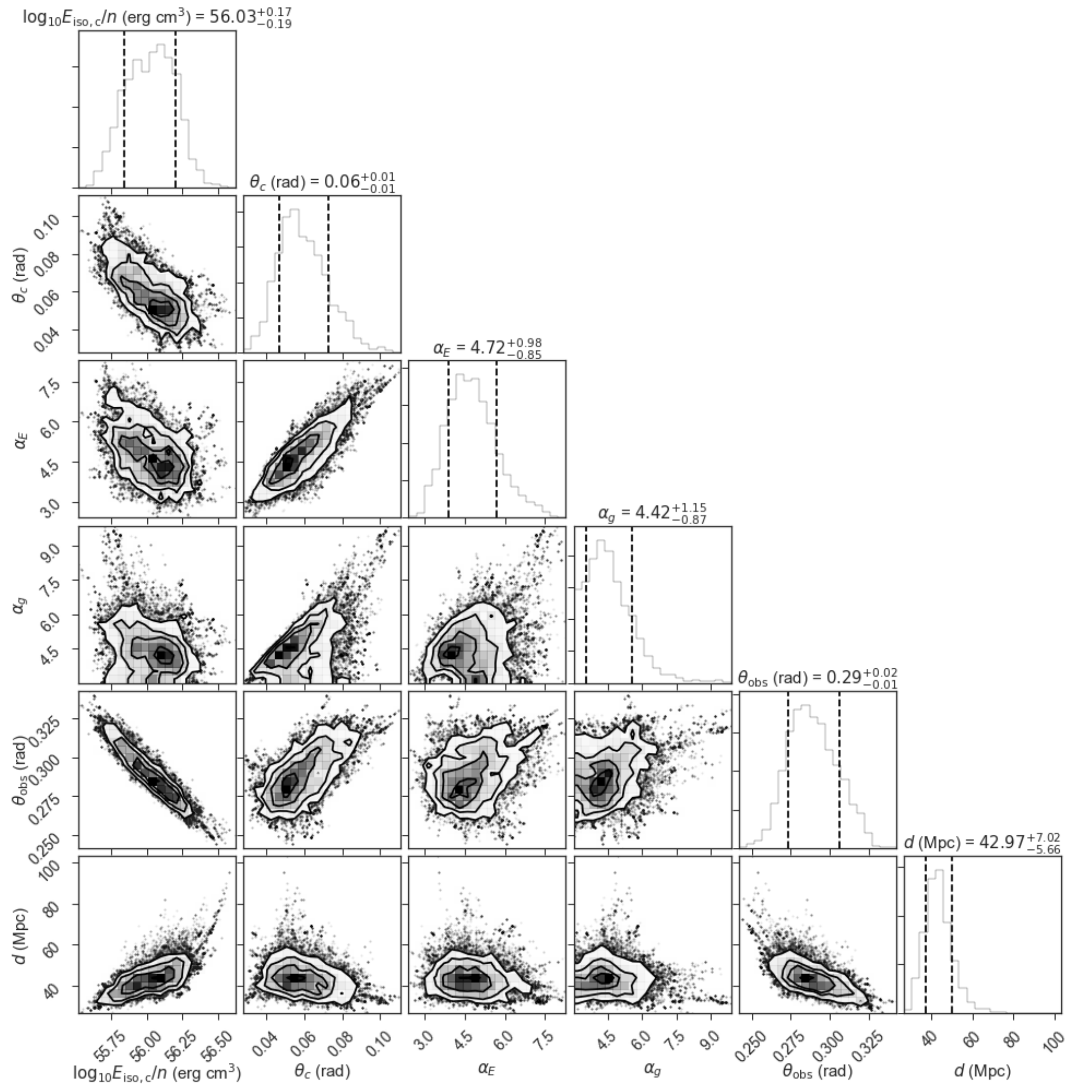}
\caption{
Corner plot\cite{corner} for a Power-Law Jet model. The afterglow light curve at $3$ GHz and the centroid motion resolved by VLBI are used as the observed input data. Vertical lines depict $68\%$ credible intervals.
}
\label{fig:PLJ}
\end{figure}

\begin{figure}
\hspace{1cm}
\includegraphics[scale=0.8]{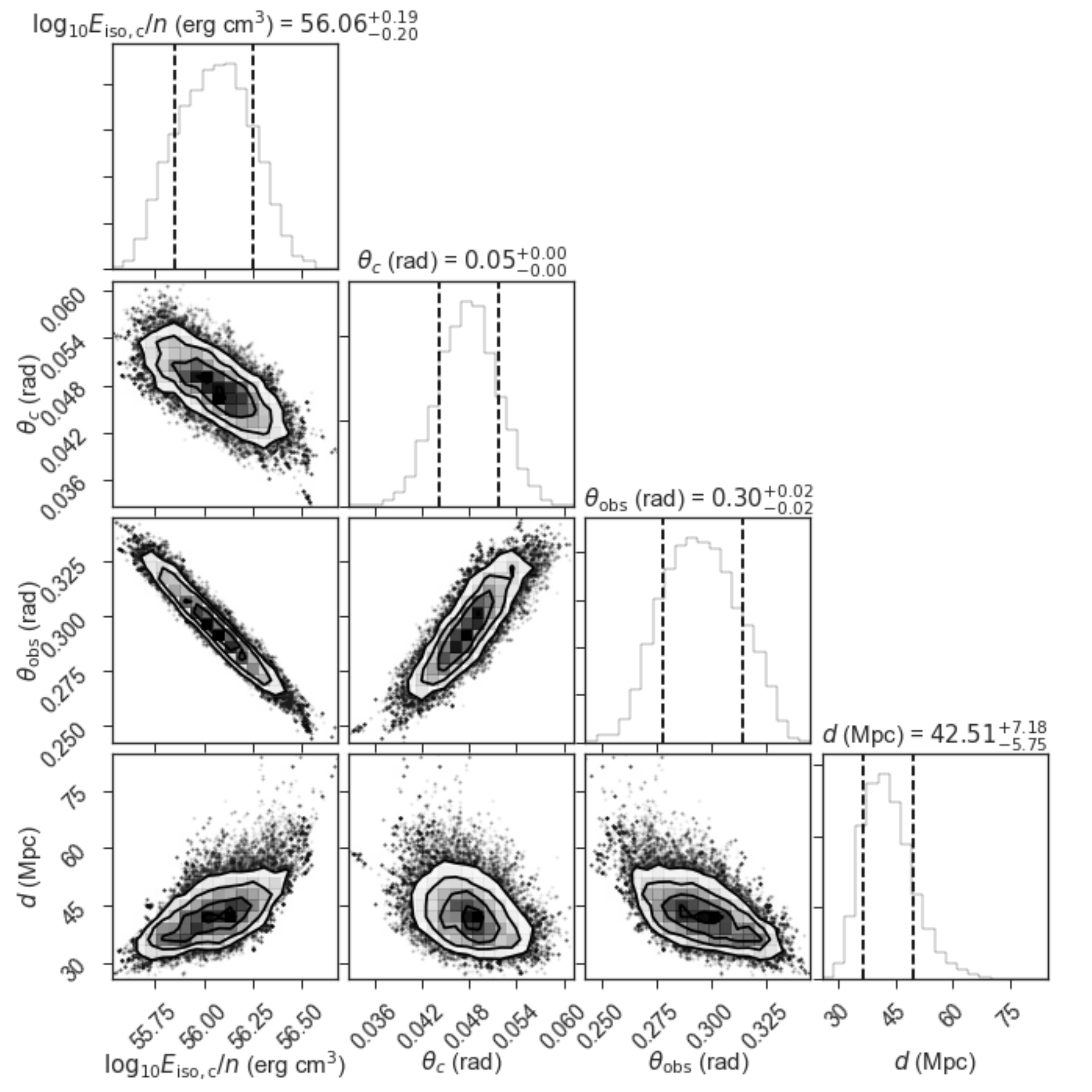}
\caption{
Same as Figure \ref{fig:PLJ} but for a Gaussian Jet model.
}
\label{fig:GJ}
\end{figure}
 
\begin{figure}
\includegraphics[scale=0.6]{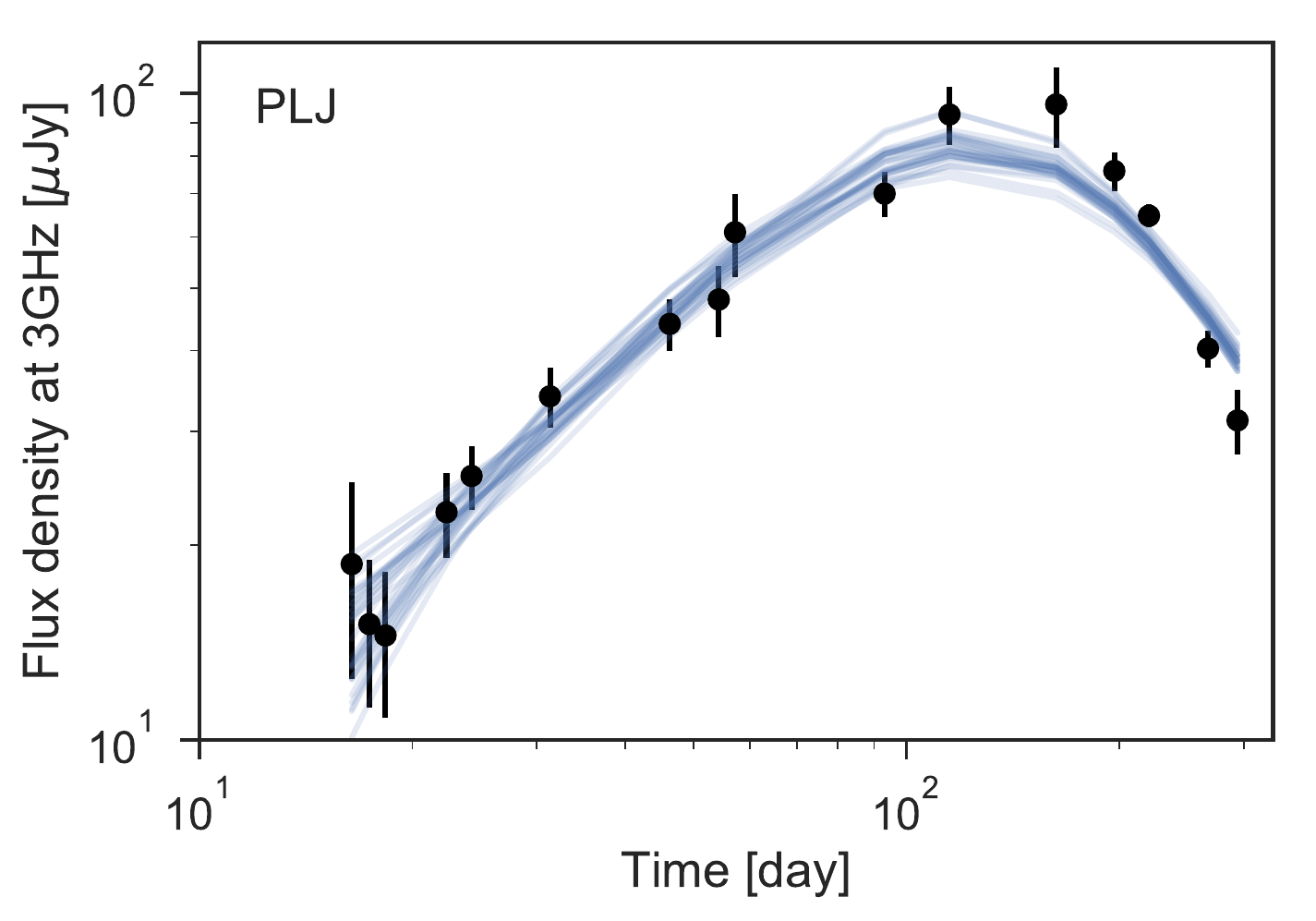}
\includegraphics[scale=0.6]{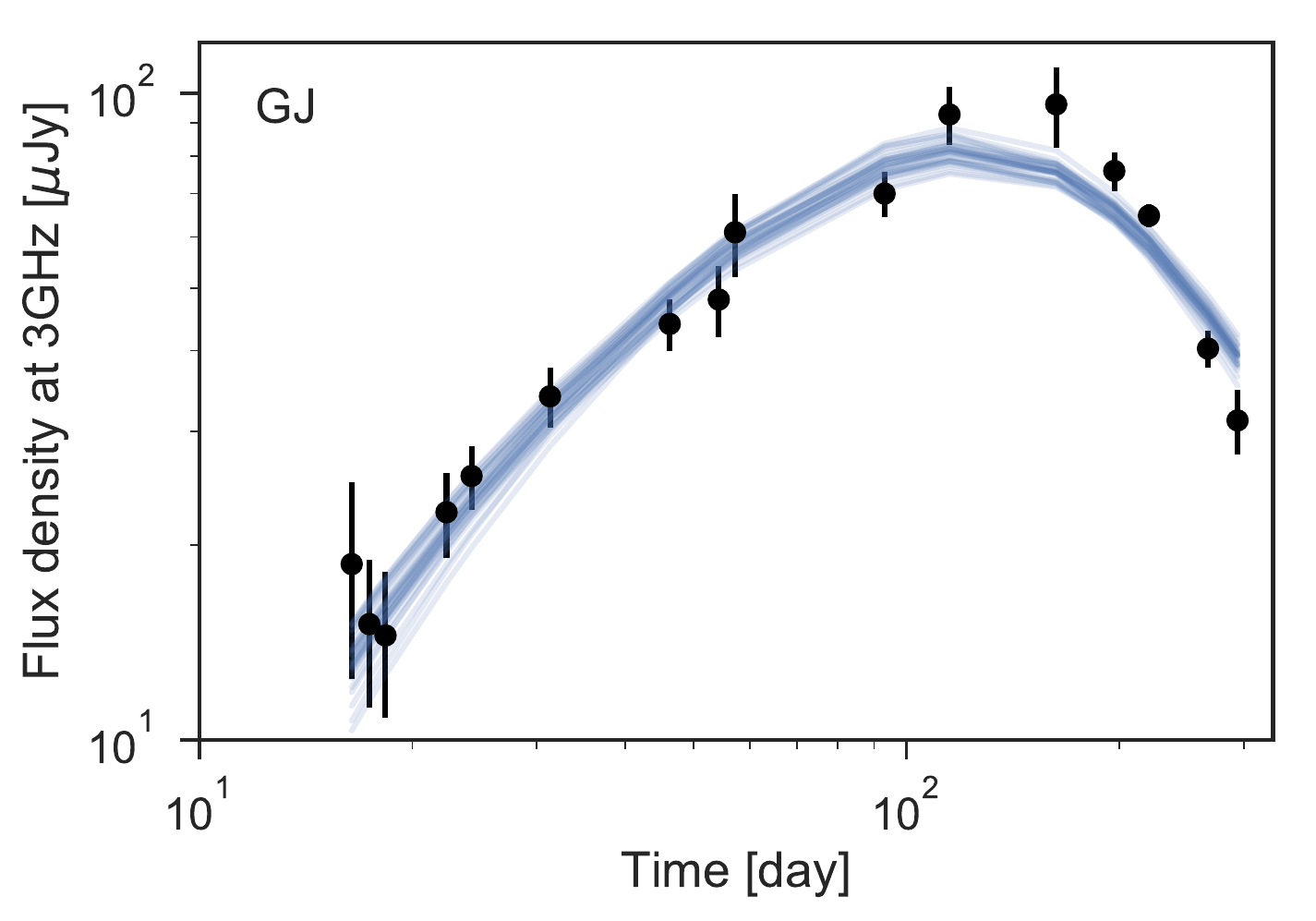}\\
\includegraphics[scale=0.6]{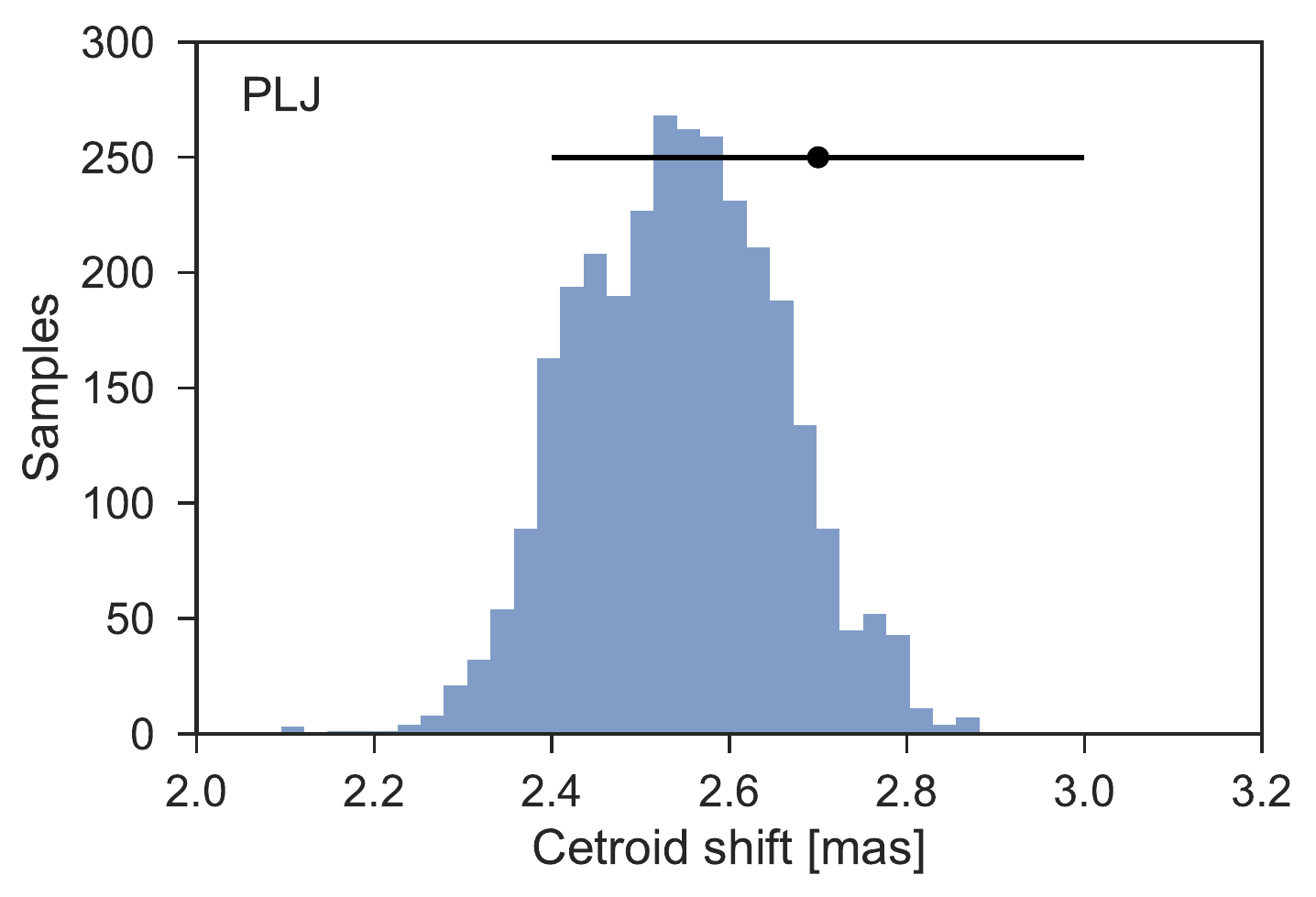}
\includegraphics[scale=0.6]{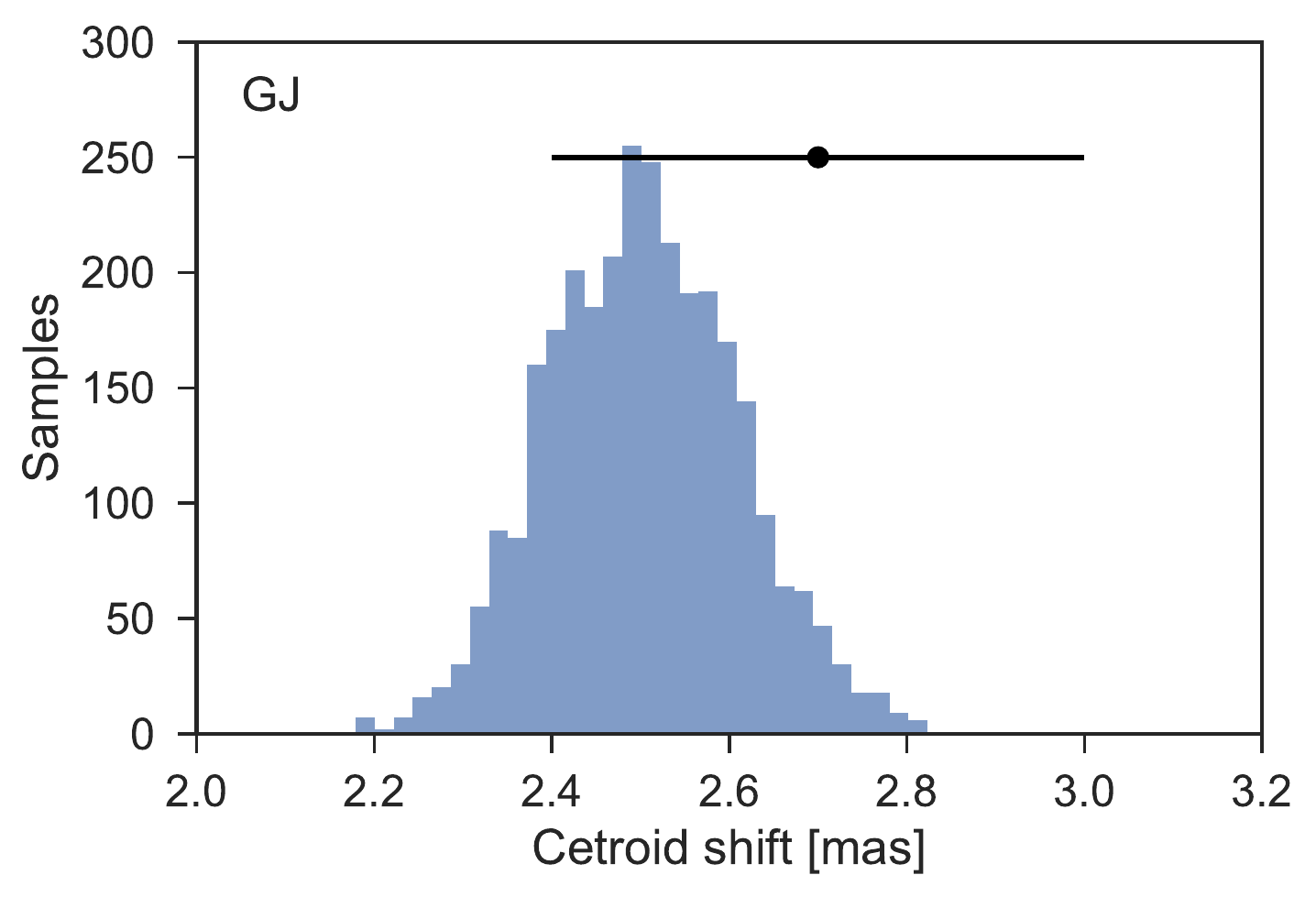}
\caption{Afterglow light curve at $3$ GHz  and centroid motion from day 75 to 230. Also shown are the light curves calculated with a PLJ ({\it upper left}) and a GJ model ({\it upper right}), where $50$ sets of the model parameters are randomly chosen from the MCMC samples.  {\it Bottom} panels show the histogram of the centroid motion with  $3000$ samples randomly chosen ({\it lower left}: a PLJ model and {\it lower right}: a GJ model). These are the results of the combined GW-VLBI-LC analysis.
}
\label{fig:exp}
\end{figure}

\begin{figure}
\hspace{0.5cm}
\includegraphics[scale=0.75]{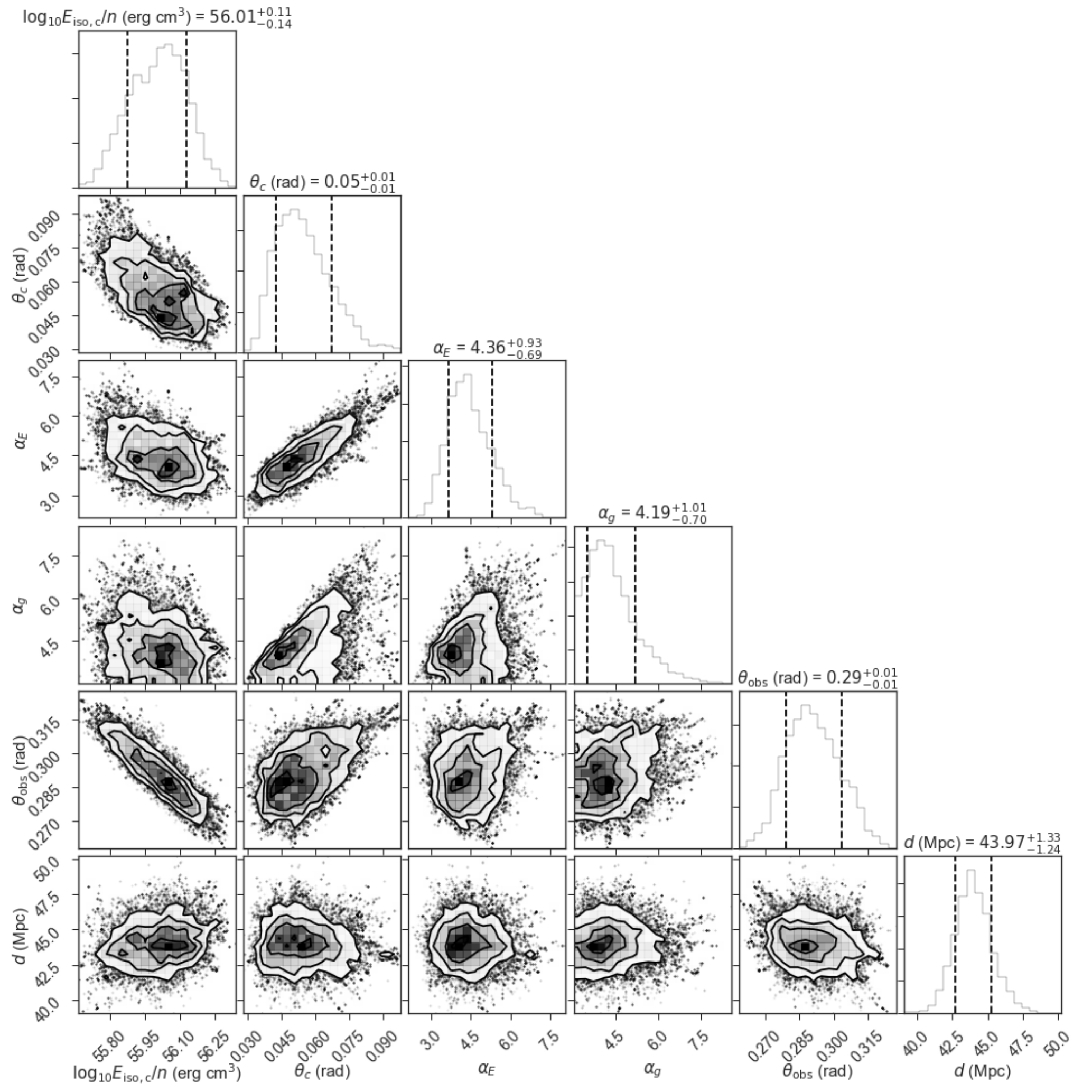}
\caption{
Corner plot for the combined GW-EM analysis with a Power-Law Jet model. The afterglow light curve at $3$ GHz and
the centroid motion resolved by VLBI are used as the observed input data. Vertical lines depict $68\%$ credible intervals. Here we use high spin PhenomNR posterior.
}
\label{fig:PLJGW}
\end{figure}

\begin{figure}
\hspace{1cm}
\includegraphics[scale=0.8]{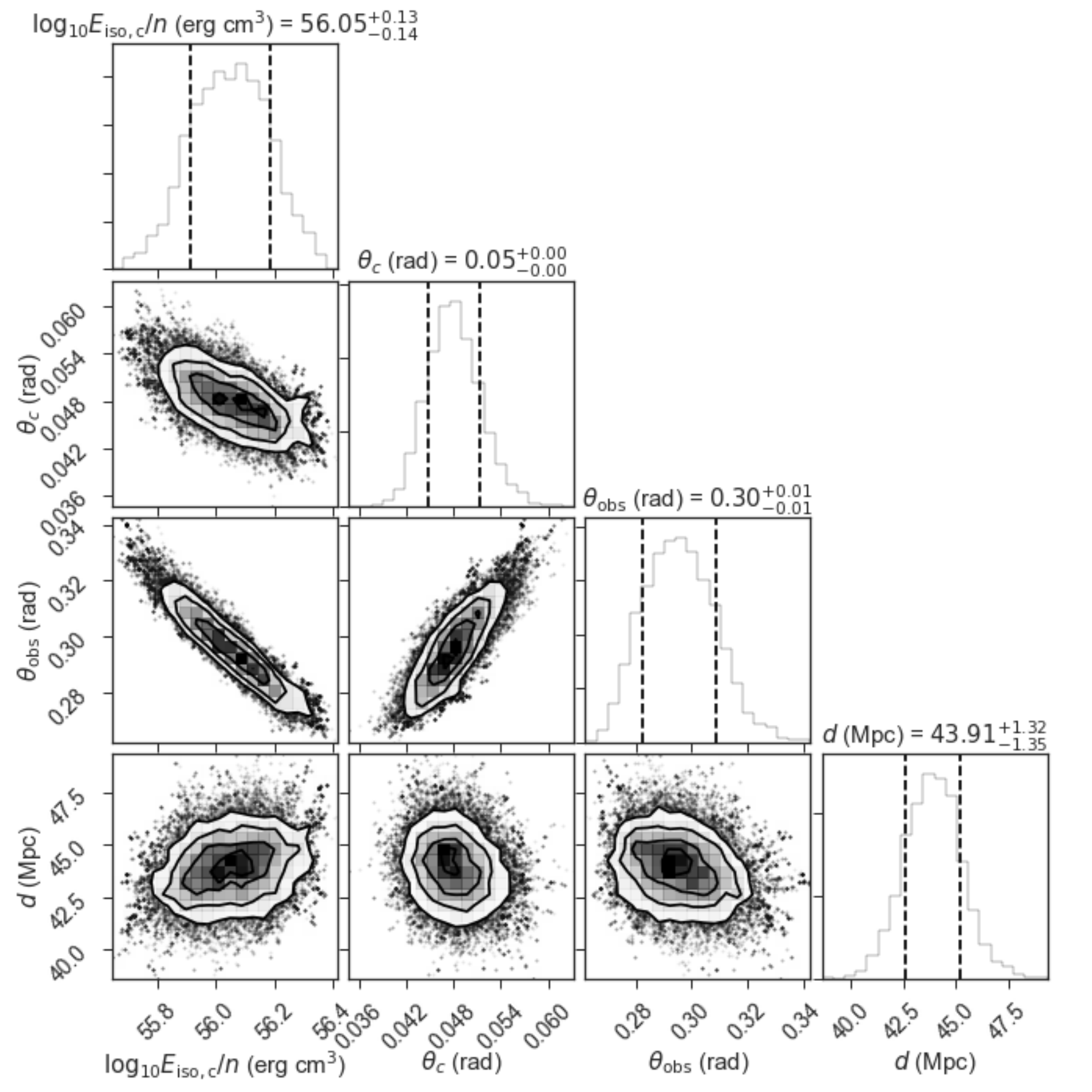}
\caption{
Same as Figure \ref{fig:PLJGW} but for a Gaussian Jet model.
}
\label{fig:GJGW}
\end{figure}

\begin{figure}
\includegraphics[scale=0.55]{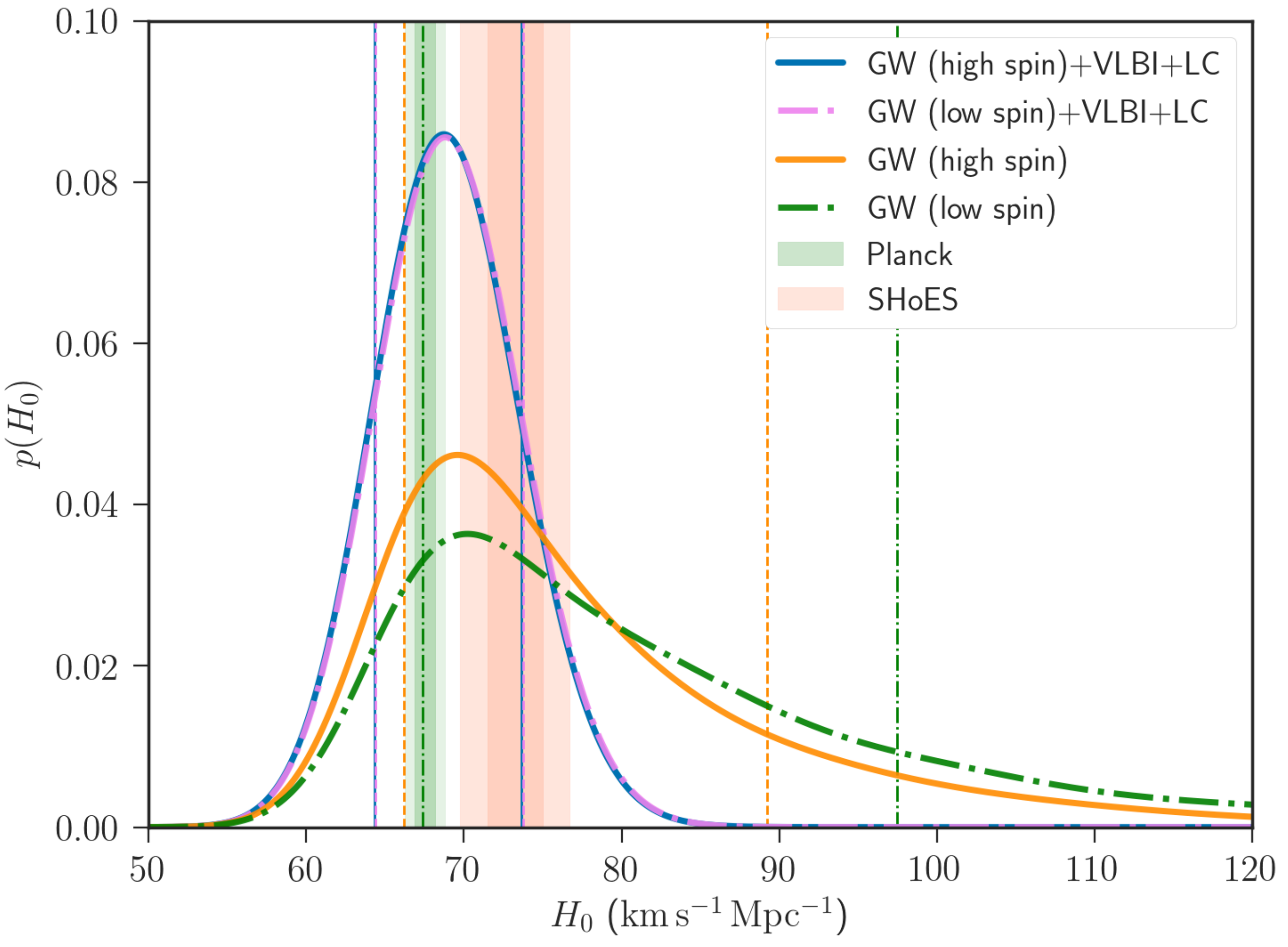}
\caption{
Comparison between the $H_0$ posteriors of the high and low spin priors.  Here we use hydrodynamics simulation jet model ($0.25<\theta_{\rm obs} \left( \frac{d}{41 {\rm~ Mpc}}\right)<0.45~ {\rm rad}$). The vertical  lines show symmetric $68\%$ credible interval for each model.  
}
\label{fig:low}
\end{figure}

\end{methods}

\clearpage

\begin{addendum}
\item The authors are grateful to Duncan Brown, Christopher Hirata, Victoria Scowcroft, Peter Shawhan, David Spergel, Hiranya Peiris for useful discussions. We thank the LIGO Scientific Collaboration and Virgo Collaboration for public access to data products. K.H. is supported by  Lyman Spitzer Jr. Fellowship at Department of Astrophysical Sciences, Princeton University. E.N. and O.G. are supported by the I-Core center of excellence of the CHE-ISF. SMN is grateful for support from NWO VIDI and TOP Grants of the Innovational Research Incentives Scheme (Vernieuwingsimpuls) financed by the Netherlands Organization for Scientific Research (NWO). The work of K.M. is supported by NASA through the Sagan Fellowship Program executed by the NASA Exoplanet Science Institute, under contract with the California Institute of Technology (Caltech)/Jet Propulsion Laboratory (JPL). GH acknowledges the support of NSF award AST-1654815. A.T.D. is the recipient  of  an Australian  Research  Council  Future  Fellowship (FT150100415).
\item[Author Contributions] K.H. carried out MCMC simulations with the synthetic models.  E.N. and O.G. derived an analytic model and carried out hydrodynamic simulation to derive constraints on the viewing angle. K.H. and K.M. analyzed the posterior samples and calculated $H_0$. G.H, K.P.M., A.T.D. provided the input observational data. K.H., E.N., S.N., G.H. wrote the paper. All coauthors discussed the results and provided comments on the manuscript.

\item[Competing Interests] The authors declare  no competing financial interests.
\item[Correspondence] Correspondence and requests for materials should be addressed to K.H. (email:\\kentah@astro.princeton.edu) and E.N. (email:udini@wise.tau.ac.il).
\item[Data Availability] MCMC samples are available from the corresponding author on request.
\item[Code Availability] The codes used for generating the synthetic light curves are currently being readied for public release. Markov chain Monte Carlo Ensemble sampler: \texttt{emcee}.
\end{addendum}


\end{document}